\documentclass[prb,aps,floats,twocolumn,twoside,preprintnumbers,
groupedaddress,floatfix,nofootinbib,showpacs]{revtex4-1}
\usepackage{amssymb,bbm,graphicx,pstricks,rotating,amsmath,gensymb}
\usepackage[T1]{fontenc}
\usepackage{slashed}
\usepackage{epsfig}

\begin{document}

\title{Embedded electron spin decoherence as indicator of the matrix material structure}

\author{Jiangyang You}
%\affiliation{Ru\dj er Bo\v{s}kovi\'{c} Institute, Division of Physical Chemistry, Bijeni\v{c}ka 54 10000 Zagreb, Croatia}
\email{Jiangyang.You@irb.hr}

\author{Dejana Cari\' c}
%\affiliation{Ru\dj er Bo\v{s}kovi\'{c} Institute, Division of Physical Chemistry, Bijeni\v{c}ka 54 10000 Zagreb, Croatia}
\email{Dejana.Caric@irb.hr}

\author{Boris Rakvin}
%\affiliation{Ru\dj er Bo\v{s}kovi\'{c} Institute, Division of Physical Chemistry, Bijeni\v{c}ka 54 10000 Zagreb, Croatia}
\email{Boris.Rakvin@irb.hr}

\author{Zoran \v Stefani\' c}
%\affiliation{Ru\dj er Bo\v{s}kovi\'{c} Institute, Division of Physical Chemistry, Bijeni\v{c}ka 54 10000 Zagreb, Croatia}
\email{Zoran.Stefanic@irb.hr}

\author{Krunoslav U\v zarevi\' c}
%\affiliation{Ru\dj er Bo\v{s}kovi\'{c} Institute, Division of Physical Chemistry, Bijeni\v{c}ka 54 10000 Zagreb, Croatia}
\email{Krunoslav.Uzarevic@irb.hr}

\author{Marina Kveder}
\email{Marina.Ilakovac.Kveder@irb.hr}
\affiliation{Ru\dj er Bo\v{s}kovi\'{c} Institute, Division of Physical Chemistry, Bijeni\v{c}ka 54 10000 Zagreb, Croatia}

%\email{josip@irb.hr}
%\affiliation{Max-Planck-Institut f\"ur Physik, (Werner-Heisenberg-Institut), F\"ohringer Ring 6, D-80805 M\"unchen, Germany}
%T\email{trampeti@mppmu.mpg.de}
%\affiliation{Institute Rudjer Bo\v{s}kovi\'{c}, Division of Theoretical Physics, Bijeni\v{c}ka 54 10000 Zagreb, Croatia}
%\email{josip@irb.hr}

%\usepackage{amsmath,amssymb,epsfig}
%\DeclareMathOperator*{\grad}{grad} \DeclareMathOperator*{\Div}{div}
%\DeclareMathOperator*{\rot}{rot} \DeclareMathOperator{\sig}{sig}
%\DeclareMathOperator{\sgn}{sgn} \DeclareMathOperator{\diag}{diag}
%\DeclareMathOperator*{\res}{res} \DeclareMathOperator{\im}{Im}
%\DeclareMathOperator{\re}{Re} \DeclareMathOperator{\id}{id}
%\newcommand{\bra}[1]{\ensuremath{\langle #1|}}
%\newcommand{\ket}[1]{\ensuremath{|#1\rangle}}
%\newcommand{\braket}[2]{\ensuremath{\langle #1|#2\rangle}}
%\newcommand{\pdiff}[2]{\frac{\partial\,#1}{\partial #2}}
%\newcommand{\pdiffn}[3]{\frac{\partial^{#3}#1}{\partial #2^{#3}}}
%\newcommand{\tdiff}[2]{\frac{d\,#1}{d #2}}
%\newcommand{\tdiffn}[3]{\frac{d^{#3}#1}{d #2^{#3}}}
%\makeatletter
%\newcommand{\fmslash}[2][0mu]{%
% \mathchoice
    %{\fmsl@sh\displaystyle{#1}{#2}}%
%    {\fmsl@sh\textstyle{#1}{#2}}%
   % {\fmsl@sh\scriptstyle{#1}{#2}}%
  %  {\fmsl@sh\scriptscriptstyle{#1}{#2}}}
%\newcommand{\fmsl@sh}[3]{%
%\m@th\ooalign{$\hfil#1\mkern#2/\hfil$\crcr$#1#3$}}
%\def\pdfannotlink{\pdfstartlink}

%\def\d#1{\partial_{#1}}
%\def\di{{\rm d}}
%\def\x{\times}
\newcommand{\tr}{\hbox{tr}}
\def\BOX{\mathord{\vbox{\hrule\hbox{\vrule\hskip 3pt\vbox{\vskip
3pt\vskip 3pt}\hskip 3pt\vrule}\hrule}\hskip 1pt}}
%\newcommand{\CC}{\mathbb{C}}
%\newcommand{\RR}{\mathbb{R}}
%\newcommand{\ZZ}{\mathbb{Z}}
%\newcommand{\ID}{\mathbb{I}}
%\def\pri{p_{r,i}}

%\renewcommand{\Im}{\mathop{\rm Im}}
%\setcounter{tocdepth}{3}

%3. Division of Theoretical Physics, Rudjer Bo\v skovi\' c Institute, P.O.Box 180, HR-10002 Zagreb, Croatia \\
%E-mail: \email{josipt@rex.irb.hr }, \email{youjiangyang@gmail.com}}

\date{\today}
%\received{\today}
%\revised{}
%\accepted{\today}

\begin{abstract}

In this work the problem of characterizing matrix material structure from embedded electron spin decoherence is studied both theoretically and experimentally. Theoretical calculation using nuclear spin bath model and cluster correlation expansion method shows that the positions of decoherence time scale extremums among single crystal orientations of the matrix material coincide with those of the nearest neighbour proton dipolar couplings. This finding is confirmed by single crystal pulsed EPR experiment performed on $\gamma$-irradiated malonic acid (MA). Electron spin decoherence decay profile in polycrystalline matrix material is obtained from the orientation dependence as an average over sampled orientations on a Fibonacci grid. In addition, it is pointed out theoretically that a further removal of crystal ordering in the nuclear spin bath can reduce decoherence time scale from the polycrystalline value. This prediction is verified experimentally by the Hahn echo time decay scale in a new amorphous polymorph of MA, obtained for the first time by mechanical milling.  Thus the embedded electron spin decoherence can be viewed as a quantitative indicator for studying structures and/or structure changes of the matrix material.

%These findings enable the view of embedded electron spin decoherence as a new type of structure indicator for the matrix material, which could have good application potential in future.

\end{abstract}

%\pacs{}

\maketitle

The electron spin phase memory/decoherence decays in Hahn spin echo and its multiple pulse generalizations, the Carr-Purcell-Meiboom-Gill (CPMG) (and, recently, the Uhrig) dynamical decoupling pulse sequences~\cite{decoupling}, are essential part of the pulsed electron paramagnetic resonance (EPR) measurements~\cite{PdEPR}.  On the other hand, the understanding of the decoherence decay itself (decay time $T_m$ and decay profile function(s)) has only taken a quantum leap forward in the last decade, when the quantum many body nuclear spin bath model~\cite{bathmodel} was developed for studying the decoherence profile of solid state quantum computer memory units (qubits)~\cite{decoherence}. In this model an isolated central electron spin's coherence is reduced by the hyperfine coupling to many ($10^2 - 10^4$) surrounding nuclear spins, which are interacting with each other via dipolar couplings. The quantum many body dynamics of the nuclear spin bath model can be solved in two dedicative methods, the linked cluster expansion (LCE)~\cite{LCE} and the cluster correlation expansion (CCE)~\cite{CCE}. LCE utilizes a perturbative expansion over the nuclear dipolar coupling and particularly suitable for semiconductor qubits. CCE, on the other hand, employs a nonperturbative nuclear spin cluster based expansion and extends the applicability of the nuclear spin bath model considerably beyond just the semiconductors.

Combined with appropriate solving methods, the nuclear spin bath model has been proven to be highly successful in predicting the decoherence behavior of various systems directly from their structure inputs~\cite{decoherence,MA,NV2,P:Si}. It then becomes natural to consider the inverse problem: How to use the nuclear spin bath model framework to connect $T_m$ back to the structure of the matrix material and, if possible, turn the experimentally long known orientational variation of  $T_m$ into a quantitative indicator for crystal structure and/or structure changes. This furnishes the main aim of this article. 

In literature it has been pointed out theoretically that in phosphorous-silicon (P:Si) qubits~\cite{P:Si}  $T_m$ can extend significantly ($> 2$ times) when magnetic field is aligned to the crystal direction [111], where the dipolar interaction between (all) nearest $\rm ^{29}Si$ nuclear spin pairs vanishes. Magnetic field orientation dependence was also used to resolve single $\rm ^{13}C$ pairs in nitrogen-vacancy (NV)-diamond system~\cite{NV1}. Two corollaries can be made from this observation: First, the maximum(s) of $T_m$ among magnetic field orientations reflect the crystal structure of the matrix material through the minimal allowable distance between nuclear spins in the bath. Second, if a structure modification can change the orientations and/or distances of the nearest nuclear spin pairs in the bath, it would influence central electron spin $T_m$ as well. 

\begin{figure}
\begin{center}
\includegraphics[width=8.25cm,angle=0]{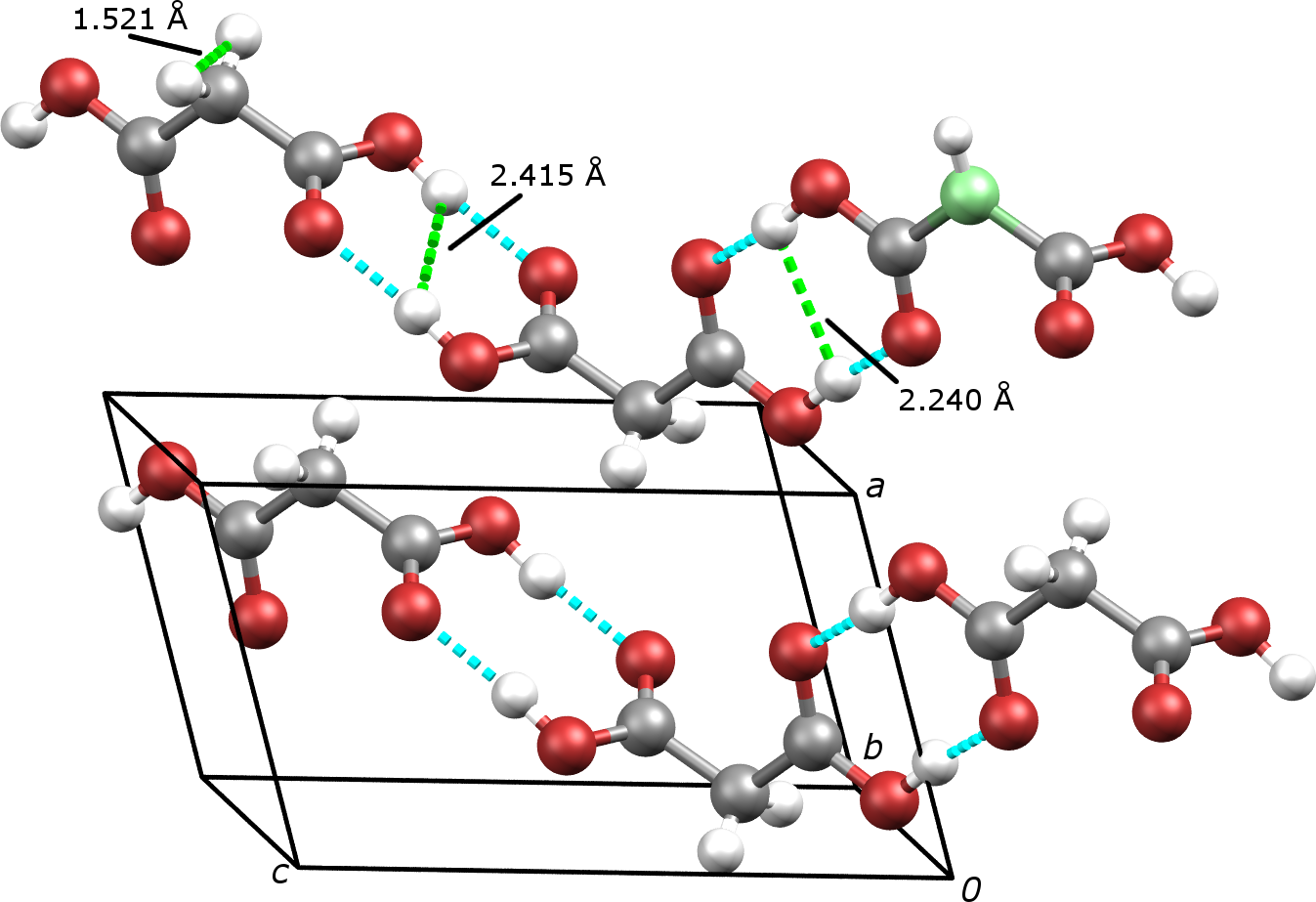}
\end{center}
\caption{Crystal structure of MA (triclinic, space group $\rm P\bar{1}$, CSD: MALNAC02)~\cite{crystal} with MA radical (next to the origin of the unit cell) embedded in:  We assume the electron density is confined on the radical carbon atom (light blue) and neglect all structure differences between the radical and molecule other than the $\alpha$-proton position. Far side shows the usual molecule chain with distances between protons in the first three closest pairs.
}
\label{MAcrystal}
\end{figure}

In this work we realize both aspects in $\gamma$-irradiated $\beta$-malonic acid (MA, Fig.~\ref{MAcrystal}), a traditional EPR benchmark material. MA is considered ideal for our study because its crystal lattice symmetry sets each of the first three nearest proton pair classes to be parallel to each other, which induces very interesting consequences as we will see below. The first corollary is observed in a single crystal rotation experiment. The second is proven by comparing the usual polycrystalline MA and a new, amorphous polymorph, obtained via mechanical milling amorphorization of polycrystalline MA.

Prior work on single crystal MA at a single orientation~\cite{MA} has shown that the nuclear spin bath provides the main contribution to the electron spin decoherence decay in Hahn echo pulse sequence.
We therefore first survey theoretically decoherence decays at various single crystal orientations using the nuclear spin bath model and CCE method. We choose to use the following spin bath Hamiltonian~\cite{bathmodel}:
\begin{gather}
H_{\rm bath}=H_{\rm Zeeman}+H_{\rm eN}+H_{\rm NN},
\\
H_{\rm Zeeman}=\mu_Bg_{zz}B_0S^z-\gamma_p\sum\limits_i B_0 I_i^z,
\end{gather}
\begin{gather}
H_{\rm eN}=\sum\limits_i E_i S^zI_i^z,
\\
H_{\rm NN}=\sum\limits_{i<j} D_{ij} \left(I_i^+I_j^-+I_i^-I_j^+-4I_i^zI_j^z\right),
\end{gather}
in which all symbols bear their usual meanings. Both the electron-nuclear spin hyperfine coupling $E_i$ and nuclear spin-nuclear spin couplings $D_{ij}$ are considered to be of the point-dipolar type, i.e.
\begin{gather}
E_i=\gamma_e\gamma_p\frac{3z_i^2-|r_i|^2}{|r_i|^5},
\\
D_{ij}=\gamma_p^2\frac{3(z_i-z_j)^2-|\vec r_i-\vec r_j|^2}{4|\vec r_i-\vec r_j|^5},
\label{Dij}
\end{gather}
where $\vec r_i$ is the the coordinate of $i$-th proton in the bath.

\begin{figure}
\begin{center}
\includegraphics[width=8.25cm,angle=0]{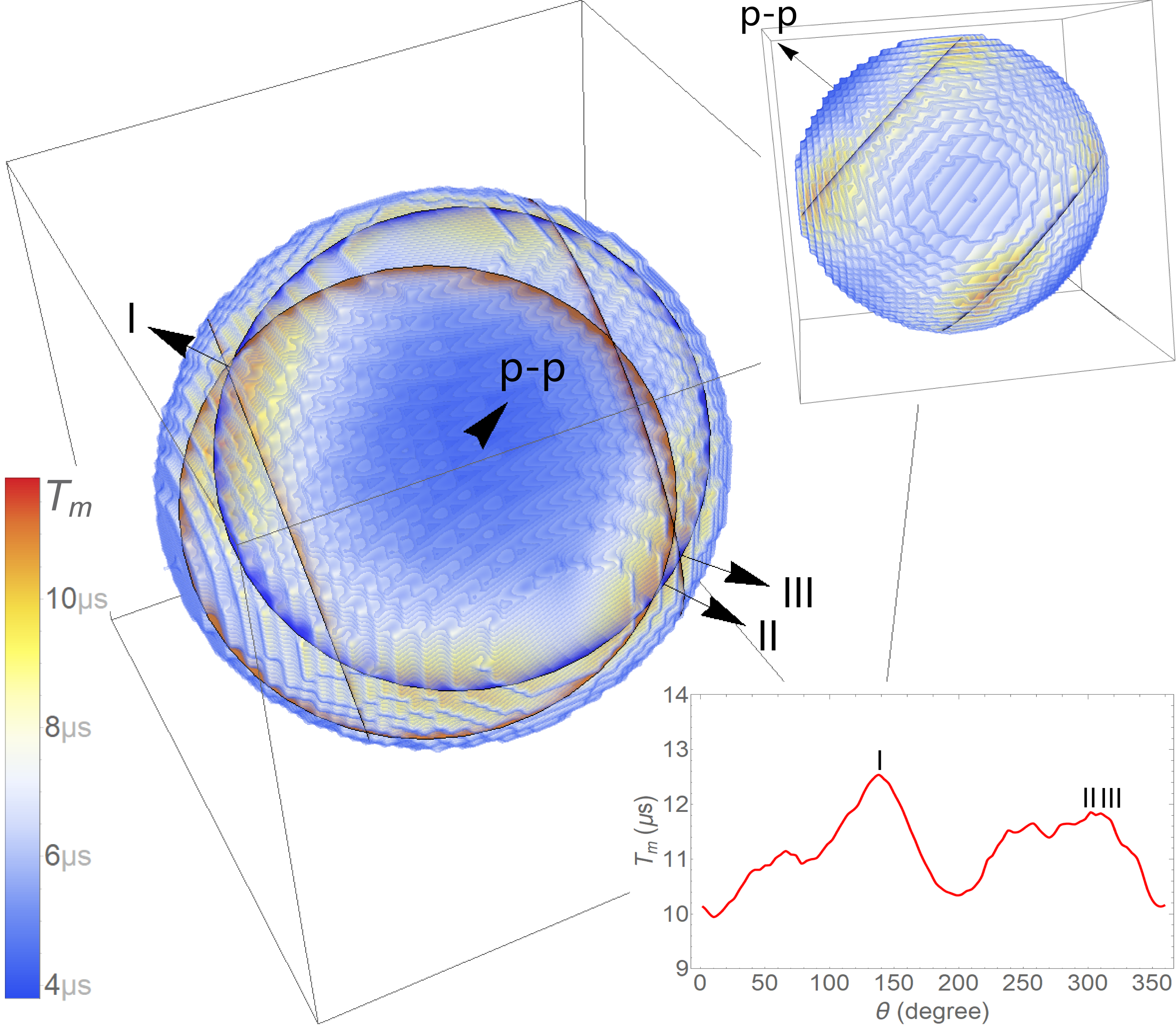}
\end{center}
\caption{Orientational distribution heat map of  $T_m$ values from nuclear spin bath model calculation. Main figure: Detailed orientational distribution of  $T_m$. The nearest proton pair direction is labelled by the ``p-p'' arrow. I, II and III mark the orientations of the calculated $T_m$ maximums on the magic angle ring. They correspond very well with the positions where the magic angle ring of the closest neighbours (blue) intersects with the magic angle rings of the second and third closest neighbours (orange). Upper insert: global distribution, two  magic angle rings (blue) can be seen passing right through the highlighted $T_m$ maximums. Lower insert: calculated $T_m$ distribution on the magic angle ring.}
\label{heatmap}
\end{figure}

We consider the nuclear spin bath consisting of one MA radical~\cite{MAradical} and 149 MA molecules, 599 protons in total. CCE till 4-clusters is caculated.  A Fibonacci grid~\cite{Fibonacci} with 251 points is used to sample the orientations. A heat map (Fig.~\ref{heatmap}) of $T_m$ values is generated by combining the Fibonacci grid and its inversion image to match MA crystal symmetry as the grid itself is not inversion invariant.
From the heatmap one can easily observe that the calculated $T_m$ values change about three times among all orientations in MA. The maximums of $T_m$ occur in the vicinity of two rings, on which  the dipolar coupling (\ref{Dij}) of all nearest pairs vanish. (Such orientations are set by the condition of being $\arccos\frac{1}{\sqrt{3}}\simeq 54.7\degree$ to the nearest proton pair direction, long known as ``magic angle'' in magnetic resonances.) The minimal $T_m$ takes place along the nearest pair direction, where dipolar coupling maximizes. These features are in accordiance to those reported in a different (P:Si qubit) system~\cite{P:Si}. Given the difference between P:Si (Fermi contact hyperfine coupling, weak nuclear spin coupling) and MA (dipolar hyperfine, strong nuclear spin coupling) systems, we consider the correspondence universal. More interestingly, the fluctuation of $T_m$ around the magic angle ring is found to be determined by the intersection with the magic angle rings of second and third closest pairs: A nearly triple intersection spot corresponds to the highest peak, while the other two slightly split ones to the pair of second highest peaks. In other words MA crystal structure actually allows all three types of closest pairs to be recorded by $T_m$ distribution.

\begin{figure}
\begin{center}
\includegraphics[width=8.25cm,angle=0]{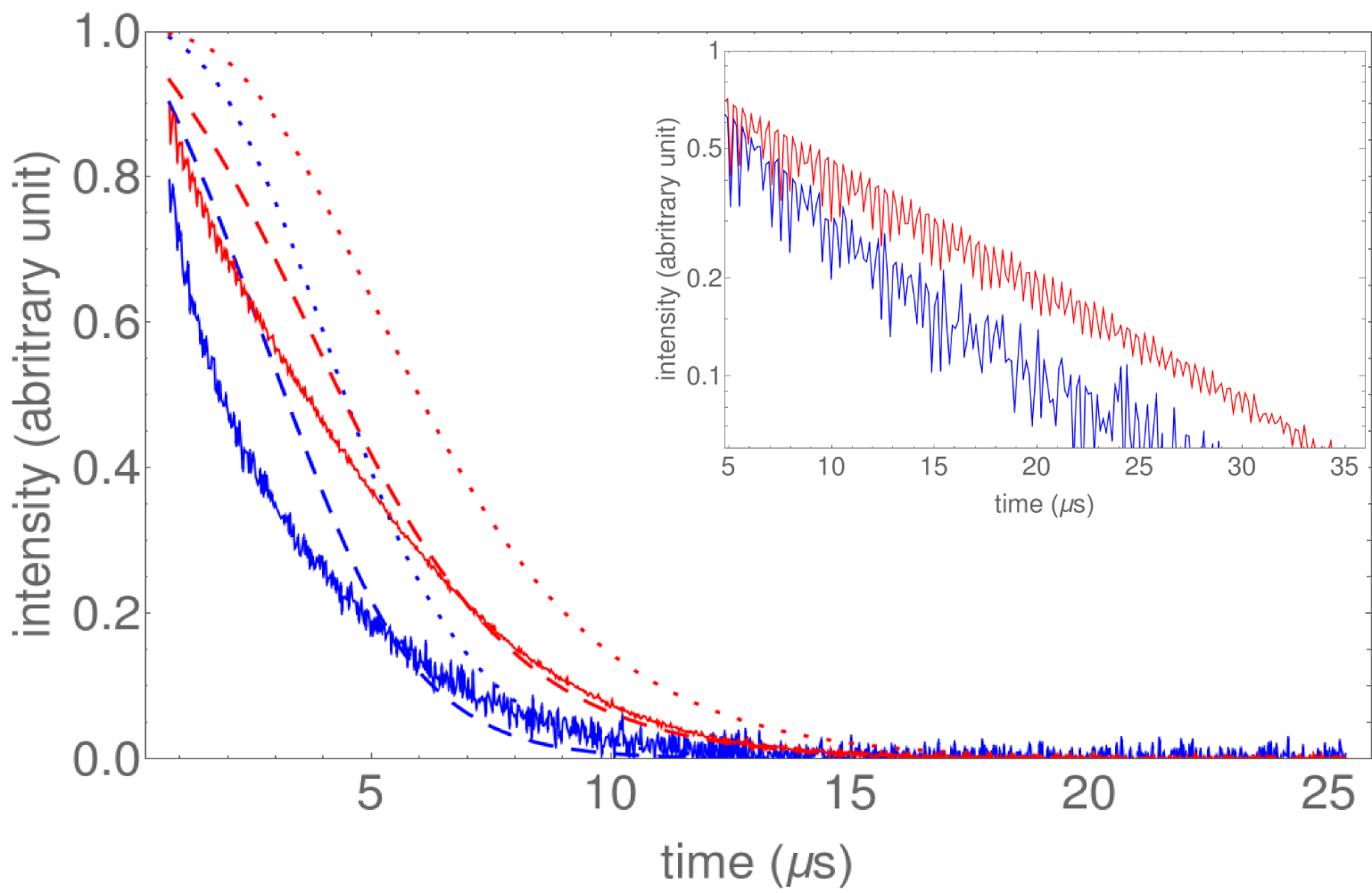}
\end{center}
\caption{Electron spin Hahn echo decay in polycrystalline (red) and amorphous (blue) MA: dotted line: nuclear spin bath model calculation result without additional correction(s); dashed line: nuclear spin bath result with (partial) corrections for electron-electron spin interaction~\cite{MA,P:Si}; solid line: experimental Hahn echo signal time decay at 50 K, normalized by an extrapolated zero time value for comparison with theory.}
\label{amorphous}
\end{figure}

By using Fibonacci grid survey we are also able to produce nuclear spin bath decay profile for the polycrystalline MA, which is obtained as a simple arithmetic average over sampling points on the grid, as each orientaion sample is approximately equally weighted in it~\cite{Fibonacci}. Approximately $50$ sampling points are found to be sufficient for consistent average. Since the polycrystalline decoherence decay profile is derived as an average over single crystal orientations, it is still controlled by the closest proton pair orientation of each. Therefore it becomes intriguing to see the consequence when MA becomes amorphous and such ordering is lost. For this purpose, decoherence decay profile of  a radical-embedded amorphous MA is estimated using a "plastic crystal" geometry, in which each MA molecule rotates randomly on its lattice position in the $\beta$-phase. Bath model calculation then shows that the electron spin coherence in amorphous MA indeed decays much faster than polycrystalline, as shown in Fig.~\ref{amorphous}.

\begin{figure*}
\begin{center}
\includegraphics[width=17cm,angle=0]{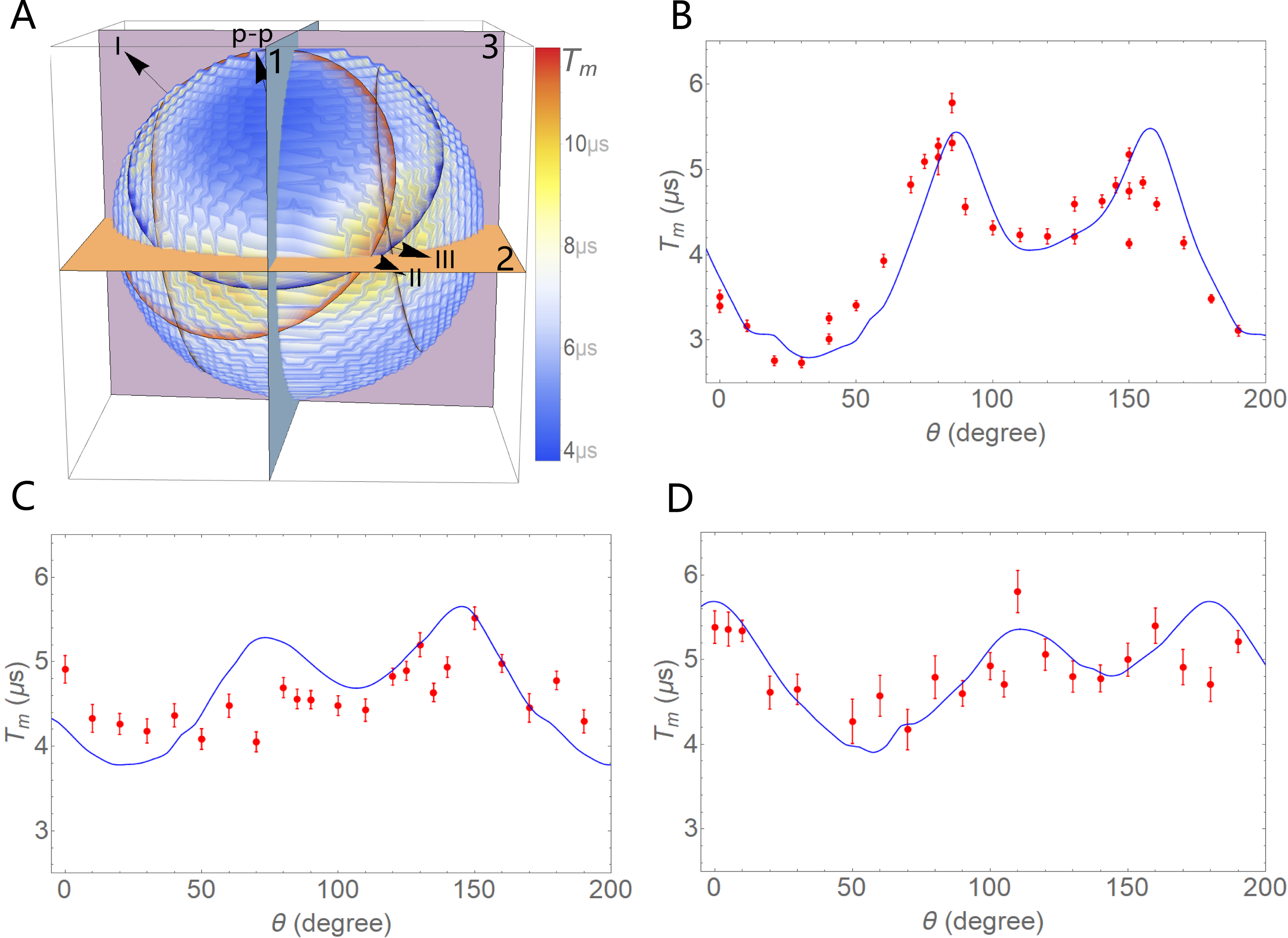}
\end{center}
\caption{
%The single crystal pulsed EPR experiment.
The comparison between single crystal $T_m$ distribution computed and measured in $\gamma$-irradiated MA crystal at 80 K. 
A: Planes of the MA crystal rotations overlaid on $T_m$ heat map; B, C, D: the theoretical (blue) and experimental (red) results of the $T_m$ distribution in crystal rotations 1, 2 and 3, respectively.} 
\label{rotation}
\end{figure*}

Based on the theory predictions the following experiments are set up as tests:

The first experiment aims at the {\em relative} distribution of $T_m$, i.e. directions of  $T_m$ maximums and minimums. In this experiment the crystal orientation dependence of $T_m$ is measured for a $\gamma$-irradiated single crystal MA at 80 K with crystal rotating around three orthogonal axes:  $\rm 1)\; a'=c^*\times b$, $\rm 2)\; c^*=a\times b$ and  $\rm 3)\; b$, as shown in Fig.~\ref{rotation}. At each orientation EPR spectrum detected via magnetic field sweep and Hahn echo signal time decay are collected.  $T_m$'s are calculated by fitting the recorded data as monoexponential decay.

In order to be compared with the experiment, theoretical bath model values of $T_m$ are rescaled by the following procedure: First, total decoherence decay profiles $f_{total}(t)$ are generated for various sampling orientations as products of the nuclear spin bath model profiles $f_{bath}(t)$ and an isotropic monoexponential decay $f_{mono}(t)=e^{-\frac{t}{\tau}}$ put by-hand:
\begin{equation}
f_{total}(t)=f_{bath}(t)\times f_{mono}(t).
\end{equation}
$\tau=7.5$ $\mu s$ is chosen to scale the theoretical $T_m$ to the experiment. Isotropy ensures that peak positions remain the same. Finally, $f_{total}(t)$'s are fitted monoexponentially to obtain rescaled $T_m$'s.

Experimental and computational results are then compared via the following independent calibration: At each sampling orientation, the resonance peak splitting is calculated using the $g$- and hyperfine tensor data from~\cite{radical}. They are then matched with the magnetic field sweep data collected in the experiment. This procedure results in a very reasonable match between theory and experiment, with only $\leq 10\degree$ miss for peak positions% as shown in Fig.~\ref{rotation}
, and thus confirms the validity of the theoretical $T_m$-orientation correspondence.

A second experiment focuses on comapring embedded electron spin decoherence in polycrystalline MA with a new polymorph which has the nearest pair alignment removed. After extensive tests it is found that MA can be amorphorized by mechanical milling with zirconium oxide ($\rm ZrO_2$) addition. (See supporting information for details.) The Hahn echo decays of electron spins in polycrystalline and amorphous MA are then studied at 50 K. Both samples are $\gamma$-irradiated to create radical concentration of $\rm 4.4\times 10^{17}cm^{-3}$ and $\rm 8\times 10^{16} cm^{-3}$ respectively according to~\cite{concentration}. Electron spin Hahn echo in amorphous sample indeed decays much faster than in polycrystalline MA despite having lower concentration. The experimental results are compared with theoretical predictions in Fig.~\ref{amorphous}. The actual decays are faster than the theoretical estimation, possibly due to the electron-electron spin interactions~\cite{Witzel}.

As a brief discussion on the major findings of this work, we would like to point out the following: 
In this work the decoherence decay profile of embedded electron spin is thoroughly proven to be closely connected with the matrix material structure. The positions of $T_m$ extremums in single crystal MA is found to coincide with those of the nearest neighbour proton dipolar couplings up to the third closest pairs, first theoretically using nuclear spin bath model and CCE method and then confirmed by single crystal pulsed EPR experiment. Such mechanism then enables a direct connection between structures of the matrix materials and $T_m$, making the latter an {\em in-situ} structure indicator for EPR experiments.

Based on the orientation dependence of  $T_m$ we predict the decoherence decay profile of $\gamma$-irradiated polycrystalline MA as a simple and expedient arithmetic average over the Fibonacci grid of sample orientations. Furthermore the nuclear spin bath model calculation using a plastic crystal geometry predicts that amorphorization of polycrystalline MA would reduce $T_m$ considerably. This effect is successfully observed experimentally by comparing Hahn echo signal time decay of $\gamma$-irradiated polycrystalline MA and amorphous MA. The latter is produced for the first time by mechanical milling.

Positive outcomes of this work convinces us to view embedded electron spin decoherence as a new type of characterization tool for matrix material structure. We expect the technique developed in this work, in particular the profiling of $T_m$ distribution by nuclear spin bath model, will be applied fruitfully in future (EPR) studies on material structures and/or structure transitions.

\begin{acknowledgments}

This work has been fully supported by Croatian Science Foundation (HRZZ) under the projects  IP-2013-11-1108,  IP-2018-01-3568, IP-2013-11-7423 and IP-2014-09-4744. The Radiation Chemistry and Dosimetry Laboratory (RCDL) at the Ru\dj er Bo\v skovi\'c Institute (Zagreb, Croatia) is gratefully acknowledged for $\gamma$-irradiating the samples. We thank Jurica Jurec, Bahar Karadeniz, Dalibor Merunka, Martina Tireli, Sre\' cko Vali\' c, Aleksandar Vi\v snjevac and Dijana \v Zili\' c for many supports, discussions and comments during this work. JY thanks Wang Yao for enlightening discussions.
\end{acknowledgments}

\end{document}